%% file: main.tex
\documentclass[runningheads,a4paper]{llncs}

\usepackage{times,amsmath,epsfig,amssymb}
\usepackage[ruled,longend]{algorithm2e}
\usepackage{graphicx}
\usepackage{caption}
\usepackage{subfig}
\usepackage{adjustbox}
\usepackage{ntheorem, lipsum}
\usepackage{url}

\newcommand{\CodeIn}[1]{{\small\texttt{#1}}}

\begin{document}
\mainmatter  
\title{Scalable Ontological Query Processing over Semantically Integrated Life Science Datasets using MapReduce}
\author{HyeongSik Kim \and Kemafor Anyanwu}
\institute{Department of Computer Science, North Carolina State University, Raleigh, NC \\
\{hkim22, kogan\}@ncsu.edu}
\maketitle
\begin{abstract} 
To address the requirement of enabling a comprehensive perspective of life-sciences data, Semantic Web technologies have been adopted for standardized representations of data and linkages between data. This has resulted in data warehouses such as UniProt, Bio2RDF, and Chem2Bio2RDF, that integrate different kinds of biological and chemical data using ontologies. Unfortunately, the ability to process queries over ontologically-integrated collections remains a challenge, particularly when data is large. The reason is that besides the traditional challenges of processing graph-structured data, complete query answering requires inferencing to explicate implicitly represented facts. 
Since traditional inferencing techniques like forward chaining are difficult to scale up, and need to be repeated each time data is updated, recent focus has been on inferencing that can be supported using database technologies via query rewriting. However, due to the richness of most biomedical ontologies relative to other domain ontologies, the queries resulting from the query rewriting technique are often more complex than existing query optimization techniques can cope with. This is particularly so when using the emerging class of cloud data processing platforms for big data processing due to some additional overhead which they introduce. In this paper, we present an approach for dealing such complex queries on big data using MapReduce, along with an evaluation on existing real-world datasets and benchmark queries.
\end{abstract}
\input{introduction.tex}
\input{preliminaries.tex}
\input{ntga.tex}
\input{evaluation.tex}
\input{relatedwork.tex}
\vspace {-1pt}
\section*{Acknowledgment}
\vspace {-1pt}
The work presented in this paper is partially funded by NSF grant IIS-1218277.
\bibliographystyle{splncs_srt}
\bibliography{IEEEexample}
\end{document}

%% file: introduction.tex
\section{Introduction}
Emerging subdisciplines in the life sciences such as Systems Biology require a more comprehensive perspective that covers broad range of data sources. This encouraged early adoption of Semantic Web technologies such as RDF and OWL\footnote{\scriptsize{http://www.w3.org/RDF / http://www.w3.org/OWL}}, as standards for representing and linking the increasing number and variety of life science databases. As a result, a number of large semantically-integrated biomedical data warehouses have emerged. For example, Uniprot~\cite{Uniprot} is a central hub for functional information on proteins, which comprises of \textit{16 sub datasets} that include core datasets like \textit{amino acid sequence}, \textit{protein names or descriptions}, \textit{taxonomic data} and \textit{citation information} as well as a lot of \textit{annotation} information from various widely accepted biological ontologies, classifications and cross-references. Another example is Chem2Bio2RDF~\cite{CHEM2BIO2RDF} which links information about interaction among chemical entities and protein molecules, i.e., integrating chemical informatics with bioinformatics within the realm of systems biology. It allows the study of the impact of small molecules towards biological systems or Chemogenomics and supports questions such as \textit{``Find genes / diseases associated with particular chemicals''}, which can be critical for pharmaceutical drug development. 

However, extracting knowledge from these data warehouses can be challenging because queries over ontologically integrated datasets requires ``inferencing'' as part of query processing to explicate entailed (implicitly represented) facts. A straightforward approach is to precompute (using forward chaining) and materialize all entailed facts to which existing pattern matching query techniques can be applied. However, besides the significant overhead of this process, it needs to be repeated whenever data is updated making it impractical for many applications. A promising direction is the use of database techniques for improving the performance and scalability of inferencing. The idea of this approach is to expand a query (rather than the data) to include all other entailed structures as alternatives as disjunctives, and then use the traditional pattern matching paradigm supported in databases to process these queries naturally. More specifically, the output of the query expansion or rewriting is a \textit{union of conjunctive queries} (UCQ)~\cite{Gottlob}. For example, suppose that we want to know all \textit{E-Coli K12} UniProt entries (including strains) and their amino acid sequences (Query 3 in Uniprot SPARQL endpoint~\cite{Uniprot}). The Uniprot \textit{core} dataset can be used to find matches for ``amino acid sequences for E-Coli K12 proteins''. However, by expanding based on the relationships in the \textit{taxonomy} dataset, all sub-taxons of the E-Coli K12, e.g., \textit{E-Coli (strain K12 / DH10B)} can also be considered. Therefore, during query processing, the pattern will be expanded to include alternatives e.g., UNION ``amino acid sequences for E-Coli (strain K12 / DH10B)'' UNION ..., etc. 

While database techniques for processing conjunctive queries, and to some degree union queries, are well-known, existing techniques are effective for queries much less complex than UCQs produced from ontological query rewriting. In particular, since the width of resulting UCQs, i.e. \#UNIONs correlates with the degree complexity of ontologies, rewritten biomedical ontological queries are significantly more complex than typical. For example, the UniProt ontology has subsumption hierarchies that are more than 25 levels deep and very broad (1,299,998 classes) in contrast to general-purpose DBPedia ontology has maximum subsumption hierarchy depth of 6 currently. Consequently, some complex UniProt ontological queries produce UCQs with over hundreds of UNION operations and thousands of JOIN operations from the expansion of \textit{Taxon} and \textit{Protein} type relations. Such queries are difficult to optimize and scale-up on large data using existing techniques. In fact, queries on datasets like UniProt are sufficiently challenging by themselves even without considering the issue of ontological query rewriting. For example, to just ask a simple query about which relationships exist between the class \textit{Protein} and \textit{Taxon} at the Uniprot SPARQL endpoint page, often produces no response at all. It appears many existing online services develop built-in optimizations for specific queries (e.g., the example queries on the website) but are unable to cope with arbitrary queries. Consequently, users may want to download the data and explore themselves using their choice of data processing platform. Since many existing ones do not deal effectively with the issue of inferencing and may be unable to cope with growing scale of these datasets (current UniProt is more than several TBs - uncompressed N-Triple format and is updated every month), a promising direction is the use of Cloud data processing platforms. Such systems support on-demand provisioning and scale-out architectures with a low barrier to entry (ease-of-use, no large upfront cost investment). However, these platforms also do not incorporate advanced enough query optimization techniques for coping with such complex queries. 

In this paper, we present an optimization approach for efficient processing of high-width UCQs on MapReduce~\cite{MapReduce} platforms. The approach is based on query rewritings that are more amenable to parallel processing both in terms of required number of computational steps and footprint of intermediate results. Specifically, our contribution can be summarized as follows:
\begin{list}{\labelitemi}{\leftmargin=0.5em}
\item An approach for efficiently evaluating complex disjunctive graph pattern queries on MapReduce platforms such as Apache Pig and Hive\footnote{\scriptsize{Apache Pig: http://pig.apache.org / Apache Hive : http://hive.apache.org}}. The approach for evaluation is based on an alternative interpretation of such queries using an algebra, the Nested Triple Data Model and Algebra (NTGA)~\cite{IEEEMagazine,ESWCRAPID,SWIMRAPID}, for modeling SPARQL graph pattern queries. 
\item Comprehensive performance evaluation demonstrating performance benefits with real life science queries and datasets.
\end{list}

%% file: preliminaries.tex
\section{Preliminaries}
\subsection{RDF/RDFS, SPARQL, and MapReduce}
An RDF model or database is a collection of statements (\textit{triples}) about resources on the Semantic Web. A \textit{triple} is a (\textit{Subject}, \textit{Property}, \textit{Object}) where \textit{Property} is a named binary relation between the \textit{Subject} and \textit{Object} (identified using URIs) or between a Subject resource and a literal value. For example, the triple (\textit{9606, commonName, ``Human''}) states that the (Taxon) 9606 has commonName ``Human'' (URI omitted for brevity). \textit{RDF Schema} (or \textit{RDFS}\footnote{\scriptsize{http://www.w3.org/TR/rdf-schema}}) includes the statements defining \textit{Classes} or Property types using a vocabulary description language written in RDF. It provides constructs to link classes (and properties, too) by subsumption relationships as well as user-defined relationships, and resources can be associated with larger collections of classes, which may be associated with typed binary relations, i.e., Properties. 

Typically, a given RDF model entails additional facts beyond those explicitly stated that can be explicated by inferencing. There are different W3C axiom sets e.g. \textit{RDFS entailment rules}\footnote{\scriptsize{http://www.w3.org/TR/rdf-mt}} for inferring implied facts in an RDF/RDFS model and RDFS. For example,  the fact that (\textit{9606 subClassOf 40674}) and (\textit{40674 subClassOf 131567}) implies that (\textit{9606 subClassOf 131567}) by transitivity of the subclass relation.  Also, \textit{domain constraint}s can be used for \textit{rdf:type} inferences, e.g., (\textit{commonName domain Taxon}) and (\textit{8801 commonName Ostrich}) implies (\textit{8801 type Taxon}). 

The standard query language for RDF data is \textit{SPARQL}\footnote{\scriptsize{http://www.w3.org/TR/rdf-sparql-query}} and its basic querying construct is called a \textit{graph pattern}. A graph pattern is a collection of \textit{triple patterns}, where each \textit{triple pattern} is a triple containing at least one variable (denoted by leading '?') in either Subject, Property or Object positions, e.g., (\textit{?taxon} commonName \textit{?name}) contains two variables. The result of the query contains mappings between datasets and variables in the query. Triple patterns sharing a common variable are interpreted as a relational join between the patterns on their variable bindings, e.g., processing the two triple patterns (\textit{?taxon} type Taxon) and (\textit{?taxon} commonName \textit{?name}) requires a single join on the subject position because of the common subject variable \textit{?taxon}.

\begin{figure}[h!]
\vspace {-12pt}
\centering
\includegraphics[scale=0.45]{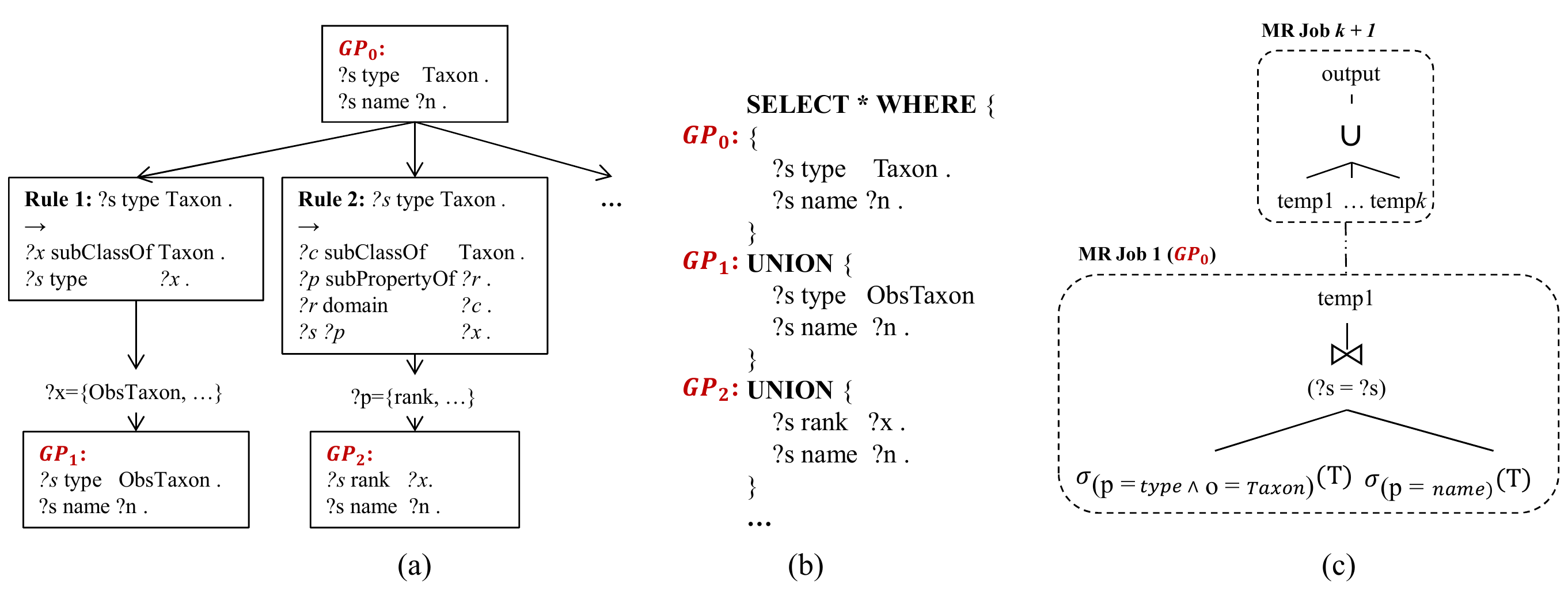}\vspace*{-5pt}
\caption{(a) Expanding original graph pattern $GP_0$, (b) resulting UCQs, and (c) the relational query plan for the UCQs.}
\label{Union}
\vspace {-15pt}
\end{figure}

To achieve scalable processing, a popular trend is to leverage a programming model called \textit{MapReduce}, which allows programs to be automatically parallelized over arbitrary-sized clusters of commodity-grade machines. Its model involves two phases \textit{Map} and \textit{Reduce}, which require expensive operations, i.e. Map: (read input, apply Map function, write the Map output) and Reduce: (shuffle/sort the Map output across nodes, apply the Reduce function, write the Reduce output). Processing complex SPARQL queries with MapReduce can be a challenge because such queries include multiple triple patterns, which require a series of joins using multiple expensive \textit{jobs} or \textit{cycles} (pairs of Map/Reduce phases). Specifically, UNION queries require a significant number of jobs, i.e. $n = k \times l + 1$ jobs where $k$ is \# of union branches, $l$ is the \# of cycles for each union clause, and 1 is the final map-only job to merge/union the output of each branch. Fig.~\ref{Union}(c) shows the relational-style execution plan that processes the example union query using $k + 1$ jobs. ($l$ is 1 because a single star pattern can be processed with a single join operation using 1 job.)

\subsection{Overview of Data Processing in RAPID+}
\begin{figure}[h!]
\vspace {-12pt}
\centering
\includegraphics[scale=0.45]{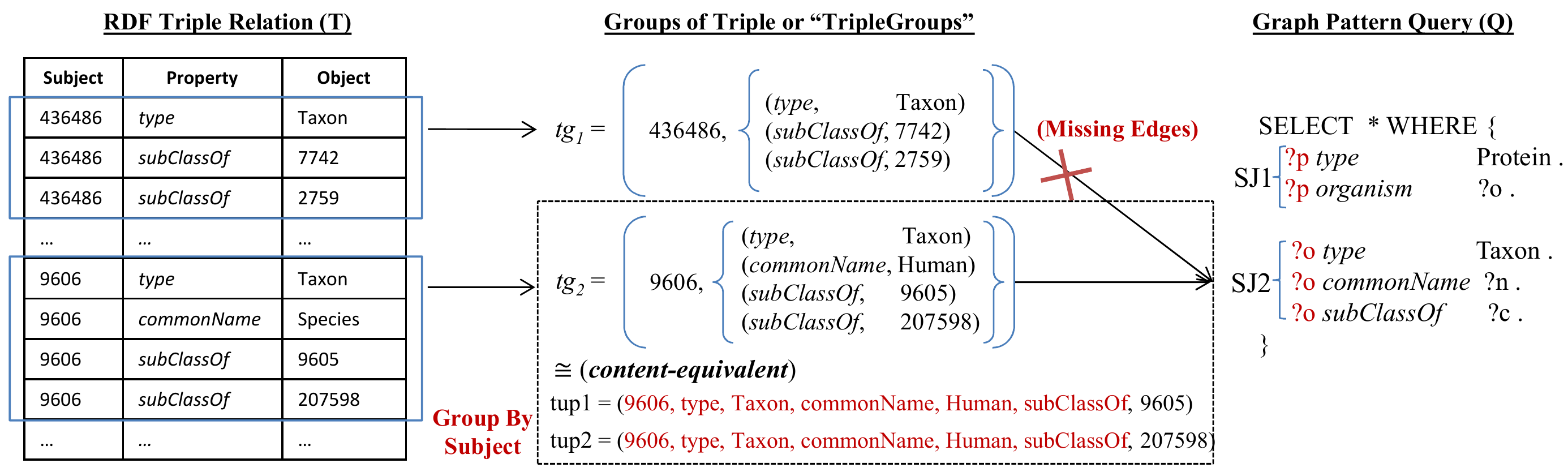}\vspace*{-3pt}
\caption {Group By on the Subject column of a triple relation $T$ produces groups of triples that are candidate matches to the star subpatterns in the query.}
\label{NTGA}
\vspace {-15pt}
\end{figure}

To overcome some of the challenges of processing complex queries in MapReduce, our earlier work in \cite{IEEEMagazine,ESWCRAPID,SWIMRAPID} proposes an alternative data model and algebra called the \textit{Nested TripleGroup Data Model and Algebra} (NTGA) for representing and manipulating RDF data. The intuition behind NTGA is to manage related sets of triples (\textit{TripleGroups}) as first-class citizens rather than the finer-grained elements, triples. Our notion of relatedness here is a group of triples having the same subject or a triplegroup. For example, $tg_1$ in Fig.~\ref{NTGA} is a \textit{Subject triplegroup} sharing common Subject \textit{436486}. Triplegroups can be typed based on the set of properties they contain, e.g., the type of $tg_1$ can be denoted as $TG_{\{\textit{type}, \textit{subClassOf}\}}$. Having triplegroups as first-class citizens goes beyond simply clustering related triples on disk, our operators manipulate and produce triplegroups rather than n-tuples. 

Two advantages of the triplegroup model are (i) conciseness of representation, similar to the advantage of the nested relational model over the classic relational model and (ii) efficiency of computation. As an example of (i), the triplegroup $tg_2$ in Fig.~\ref{NTGA} represents the same information as, or is  \textit{content-equivalent} to the set of n-tuples \{\textit{tup1}, \textit{tup2}\} (a notion stated more precisely in earlier work). Intuitively, we can produce an equivalent set of n-tuples from a triplegroup by splitting the triples in the triplegroup based on the property types, and then applying a Cartesian product between the resulting subsets, e.g., $tg.triples(type)$ $\Join$ $tg.triples(commonName)$ $\Join$ $tg.triples(subClassOf)$ where $tg.triples(prop)$ is the set of triples in $tg$ with property type $prop$. More precisely, since triples in a triplegroup all share the same subject, the content-equivalence relationship is always with the set of n-tuples resulting from a star-join. With respect to (ii), it is possible to compute all triplegroups represented in data (irrespective of structure) using an operation similar to a relational GROUP BY operation on the Subject column. This is the semantics of the NTGA operator \CodeIn{TG\_GroupBy($\gamma$)}. The triplegroups generated from this operator can be considered as valid matches for multiple star subpatterns or star-joins in a query if they contain all the properties in the triple patterns of the star subpattern. For example, the triplegroups $tg_2$ in Fig.~\ref{NTGA} can be seen as valid matches to the star subpatterns $SJ2$ because each of them contains all the required properties listed in each star pattern. However, $tg_1$ does not match any sub star patterns listed in the query since it does not include triples with some of the properties required by the star patterns, e.g., \textit{commonName} for $SJ2$. The operator called \CodeIn{TG\_GroupFilter($\gamma^{\sigma}$)} filters out such unmatched triplegroups, i.e. triplegroups that do not contain ``matching'' subsets to \textit{any} of the star subpatterns of the query $Q$. Note that \CodeIn{TG\_GroupBy} and \CodeIn{TG\_GroupFilter} use only 1 job to produce all valid matches for all the star subpatterns whereas relational-style approach has to use a separate job for each star subpattern to produce the equivalent output.

\begin{figure}[h!]
\vspace {-12pt}
\centering
\includegraphics[scale=0.44]{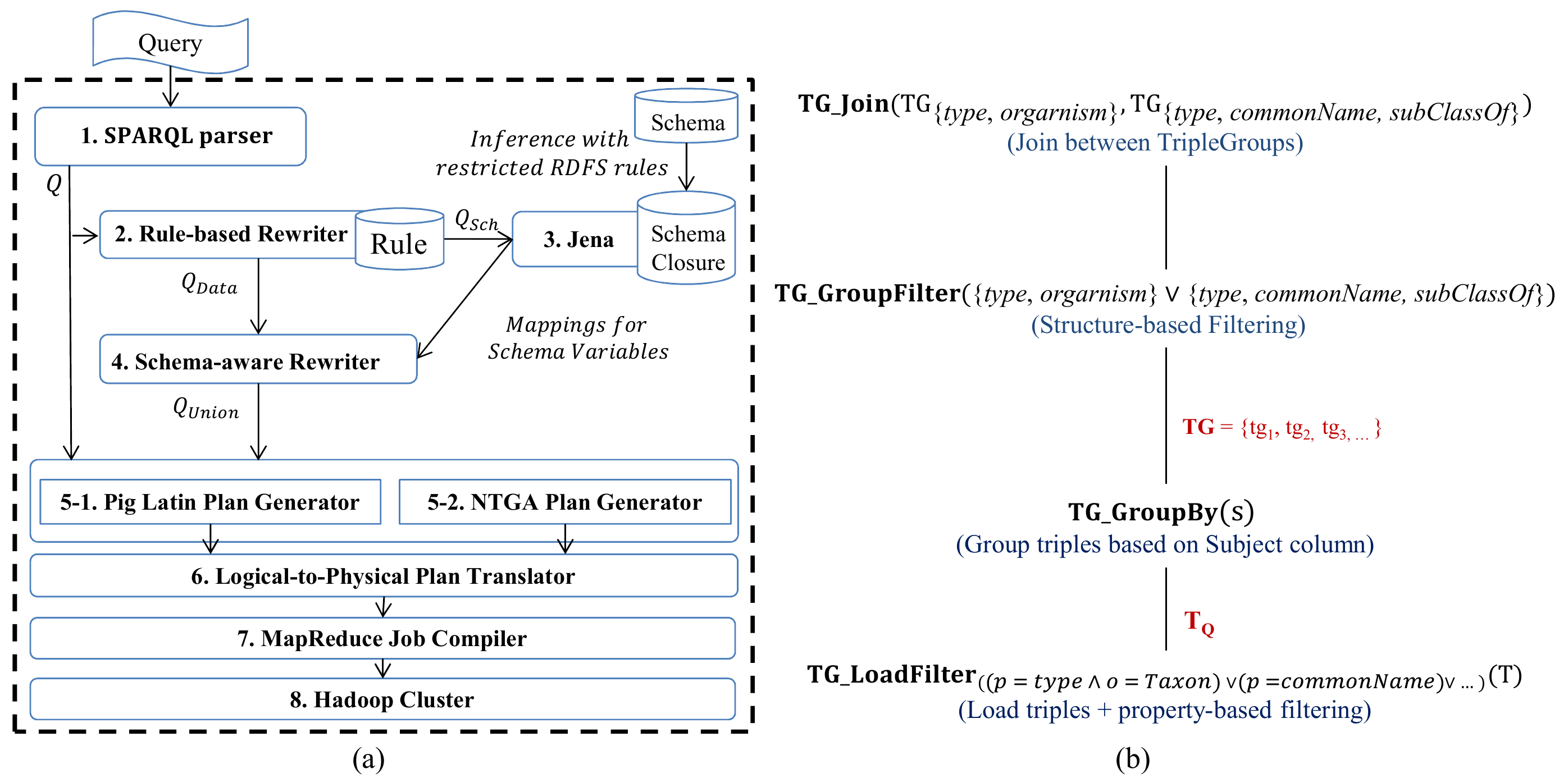}\vspace*{-3pt}
\caption {(a) An overall RAPID+ system architecture, (b) The NTGA-based logical plan for the query in Fig.~\ref{NTGA}, and (c) The NTGA-based MapReduce execution plan for the union query in Fig.~\ref{Union}(b).}
\label{Architecture}
\vspace {-15pt}
\end{figure}

Fig.~\ref{Architecture}(b) shows the NTGA-based logical plan for the query $Q$ in Fig.~\ref{NTGA}(b). The mapping of this logical plan to a MR execution workflow highlights its advantages on MapReduce. \CodeIn{TG\_LoadFilter}, \CodeIn{TG\_GroupBy}, and \CodeIn{TG\_GroupFilter} are mapped to the first job, which filter out unnecessary triples and construct star-subgraphs (triplegroups) relevant to the query. Additional jobs are then required to join star-subgraphs using TG\_Join($\gamma^{\Join}$) if the query includes multiple star patterns. The maximum number of additional cycles will be $n$, where $n$ is the \# of star pattern subqueries plus the first job. In contrast, using the relational-style interpretation of the queries requires $(2n-1)$ jobs. The shorter execution workflow length leads to savings and benefits given that each job incurs costs of multiple disk and network I/Os. More detailed explanations and algorithms of the operators are available in \cite{ESWCRAPID}.

%% file: ntga.tex
\section{Scalable UCQ Processing using RAPID+}
We begin with an overview of the architecture of \textit{RAPID+} which is an extension of Apache Pig that we have developed for scalable and efficient processing of Semantic Web workloads using NTGA. We then discuss the generation and the execution of UCQs in RAPID+.
\subsection{UCQ Generation in RAPID+}
RAPID+ includes the custom data structures for managing triplegroups and components for supporting execution of NTGA-based query plans. Fig.~\ref{Architecture}(a) shows the general process flow of a graph pattern query $Q$ in RAPID+. The query is first parsed via an integrated SPARQL query interface using Jena's ARQ or an extended Pig Latin interface. This is then fed into the \textit{Plan Generator} translating the query expression tree to i) the logical plan represented with the operators in Pig or ii) the plan represented with NTGA operators. The logical plan is then fed into the \textit{Logical-to-Physical Plan Translator}, and compiled by the \textit{Job Compiler} in Pig, producing a workflow of MR jobs. Additional details on the architecture of the system and the process flow can be found in \cite{IEEEMagazine}. In this paper, three new components have been added to generate UCQs between query parsers and plan generation layers. Once parsing the query is done, our \textit{Rule-based Query Rewriter} in Fig.~\ref{Architecture} expands the query, producing the initial UCQs. The schema triple patterns in each union branch ($Q_{Sch}$) are then executed on the Jena locally first, generating the variable mappings for the schema variables. With these mappings, the \textit{Schema-aware Rewriter} rewrites the remaining part of the union branches ($Q_{Data}$), generating the final UCQs ($Q_{Union}$), e.g., the example query in Fig.~\ref{Union}(a). 

\begin{figure}[h!]
\vspace {-12pt}
\centering
\includegraphics[scale=0.5]{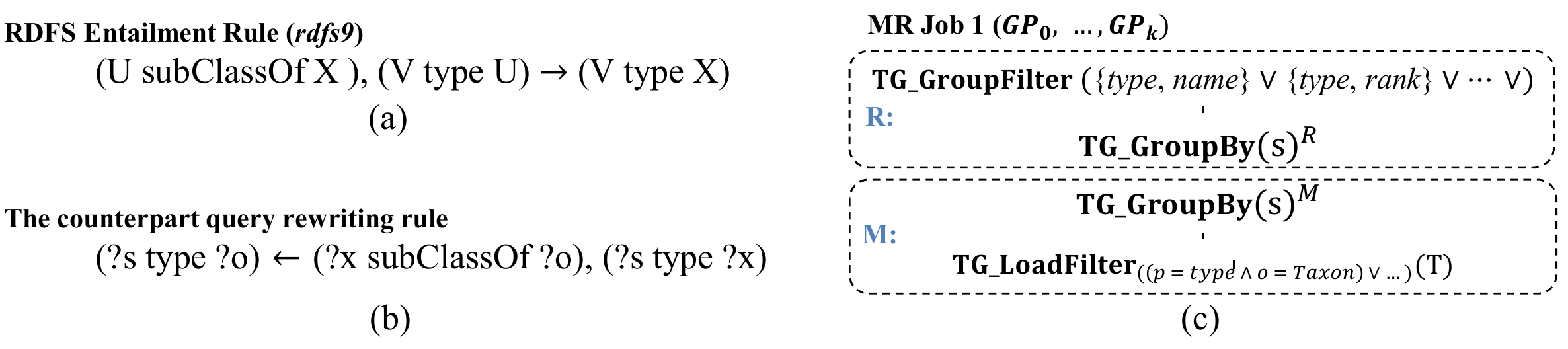}\vspace*{-3pt}
\caption {(a) The RDFS entailment rule (\textit{rdfs9}), (b) the corresponding query rewriting rule, and (c) The NTGA-based MR job for the UCQs in Fig.~\ref{Union}(b) (\textit{M} and \textit{R} denote Map and Reduce).}
\label{UCQ}
\vspace {-15pt}
\end{figure}

We extend the hybrid approach in \cite{TradeOff} to generate UCQs for RDFS entailment. In this approach, the partial closure is computed offline using the restricted entailment rules that avoid deriving the triples with the property \textit{rdf:type} (or \textit{type} triples). The \textit{rdf:type} inferences are later performed online by rewriting the \textit{type} triple patterns into UCQs and executing the UCQs. In this paper, we adapt the approach to generate the partial closure only for the ontology schema, called the \textit{Schema Closure}. In addition, we rewrite the generated UCQs at runtime as follows.
\vspace {-4pt}
\begin{enumerate}
\item for each union branch, find the mappings between the schema closure and the variables in the schema triple patterns, which contain the properties predefined in RDF Schema. 
\item remove the schema triple patterns, rewrite other remaining triple patterns using the mappings, and then execute the re-written UCQs on MapReduce.
\vspace {-4pt}
\end{enumerate}
For example, the box \textit{Rule 1} of Fig.~\ref{Union}(a) shows that the query rewriting rule in Fig.~\ref{UCQ}(b) is applied as the counterpart of the rule (\textit{rdfs9}) in Fig.~\ref{UCQ}(a), which derives new type statements using \textit{subClassOf} relationship. We then retrieve all subclass statements from the schema closure by executing only the first schema triple pattern with the property \textit{subClassOf}, i.e, only execute the pattern (\textit{?x subClassOf Taxon}) in the box. The mappings for the variable $?x$ are then retrieved as \textit{x = \{ObsTaxon, ... \}} (The subclasses of \textit{Taxon}). Once these mappings are generated, we remove the schema triple patterns and substitute the variable $?x$ with the mappings, e.g., Fig.~\ref{Union}(a) shows that the original pattern (\textit{?s type ?x}) in $GP_0$ is now rewritten as (\textit{?s type ObsTaxon}) in $GP_1$. For the entailment rules related to other properties such as \textit{rdfs:domain}, \textit{rdfs:range} and \textit{rdfs:subPropertyOf} (i.e., \textit{rdfs2}, \textit{rdfs3}, and \textit{rdfs7}), we similarly extend the graph patterns using the corresponding rewriting rules, producing new graph patterns shown in Fig.~\ref{Union}(b).
\subsection{UCQ Execution in RAPID+}
The remaining issue is then how UCQs can be processed using NTGA. We first consider the case in which all UNION subqueries contain a single star pattern as the only graph pattern, i.e. a graph pattern $GP_1$ in a union branch only contains a single star sub pattern $SJ_1$. Intuitively, NTGA optimizes this case by transforming the union of joins into the logical disjunction of joins by applying the splitting law backwards, i.e. the two or more selections can be merged into the selections involving a logical OR ($\sigma_{C_1}(R)\;\cup_{Set}\;\sigma_{C_2}(R) = \sigma_{(C_1\;\vee\;C_2)}(R)$ where $C_1$ and $C_2$ are the conditions of the selection operators($\sigma$)). In \CodeIn{TG\_GroupFilter}($\gamma^{\sigma}$) operator, multiple star subpatterns can be evaluated concurrently in a single data pass by including them with logical ORs as the parameter of the operator, e.g., $\gamma^{\sigma}(SJ_1 \vee SJ_2$) if the graph pattern $GP_1$ consists of the two star patterns $SJ_1$ and $SJ_2$. Leveraging this ability, the graph patterns ($GP_1, GP_2, ...$) in union branches connected by logical OR as the parameter can be paramaterized, i.e. transforming $\gamma^{\sigma}(GP_1)\cup_{Set}\gamma^{\sigma}(GP_2)$ to $\gamma^{\sigma}(GP_1 \vee GP_2)$. This is the scenario shown in Fig.~\ref{Union}(b) where the example query contains $k$ branch of unions which can be computed in a single pass. While the relational-style plan process this query with $k+1$ jobs, only 1 job is required with NTGA as shown in Fig.~\ref{Architecture}(c). 

For the case where each union branch contains multiple star patterns, NTGA requires additional join operations to connect stars in each union branch. Instead of processing them using separate jobs, NTGA first traverses the query graph to find and group common star structures and join variables in different union branches. Once such stars are discovered, NTGA processes the joins between star patterns together. For this purpose, we introduce an extended \CodeIn{TG\_Join} that can manage the joining with different classes of join patterns called \CodeIn{TG\_UJoin}. This operator can be considered as a variant of the \CodeIn{TG\_Join} that can accept more than one star patterns as a left/right operand using primitives similar to a logical OR. The plan using this operator will use the lesser number of jobs for join operations, which eventually shortens the overall execution time.

%% file: evaluation.tex
\section{Evaluation}
\subsection{Setup}
\textbf{Cluster Configuration:} The experiments were conducted on a 80-node Hadoop\footnote{\scriptsize{Apache Hadoop (http://hadoop.apache.org) is the open-source implementation of the MapReduce framework.}} cluster in VCL\footnote{\scriptsize{Virtual Computing Lab: http://vcl.ncsu.edu}} where each node was equipped with a dual core x86 CPU (2.33 GHz) and 4GB RAM. All results recorded were averaged over three or more trials. 

\noindent \textbf{Datasets:} Our testbed consists of two life science datasets: Uniprot~\cite{Uniprot} (May 2013) and Chem2bio2RDF~\cite{CHEM2BIO2RDF}. The core dataset in Uniprot includes approx. 6.8B triples (968GB in N-triple). Chem2bio2RDF consists of 25 sub datasets extracted from various sources, e.g., \textit{PubChem}\footnote{\scriptsize{http://pubchem.ncbi.nlm.nih.gov}} and contains approximately 412M triples (71GB in N-triple).

\noindent \textbf{Techniques:} Hive 0.10.0 was selected as a benchmark target with the two types of the plans: the plan with UNION operators (Hive(Union)), and the plan using the multi-query optimization (MQO) technique in \cite{MQO} (Hive(Optional)). For Hive(Optional), we adapt the technique to merge common subpatterns across different UNION branches, which produces the query that include the common subexpression as a root graph pattern and the rest patterns in OPTIONAL clauses. This allows the common sub expression to be executed once and shared across the branches. For some cases, the MQO was not applied, e.g., \CodeIn{UQ3} does not contain any common sub expressions. 

\noindent \textbf{Test Queries:} The evaluation involves queries without and with a UNION operator (either explicitly or by rewriting for inference). The former was included as a baseline comparison between relational vs. NTGA execution plans. All queries were selected from the lists on the websites for Uniprot\footnote{\scriptsize{http://beta.sparql.uniprot.org}} and Chem2bio2RDF\footnote{\scriptsize{http://chem2bio2rdf.wikispaces.com}}, respectively. 9 out of 18 queries were selected from the Uniprot queries (others included constructs such as path expressions that are not currently supported in NTGA).  For similar reasons, 5 queries were selected from the list of example Chem2bio2RDF queries. Table \ref{QueryDetails} gives the characteristics of queries used, e.g., the number of the triple patterns and the star patterns (\#TP and \#STP), the number of the edges in each star patterns (\#Edges in STP) denoted as colon separated list e.g. X:Y:Z implies 3 star patterns with X, Y and Z edges respectively, the number of the subject-object/object-object joins (\#S-O and \#O-O), the width of the UNION operator (\#Br in $\cup$). Additional details about the evaluated queries and the results are available on the project website.\footnote{\scriptsize{http://research.csc.ncsu.edu/coul/RAPID+/CSHALS2014.html}} 
    \begin{table}[!ht]
    \vspace {-14pt}
    \centering
    \footnotesize
    \adjustbox{scale=0.75}{
    \begin{tabular}{| p{1.1cm} | p{0.72cm} | p{0.72cm} | p{1.31cm} | p{0.81cm} | p{0.81cm} |}
    \hline
	\textbf{Query (w/o $\cup$)} & \textbf{\#TP} & \textbf{\#STP} & \textbf{\#Edges in STP} & \textbf{\#S-O $\Join$} & \textbf{\#O-O $\Join$} \\ \hline		
	UQ1	& 1	& 1	& 1 & 0	& 0 \\ \hline
	UQ2 & 3 & 1	& 3	& 0 & 0 \\ \hline
	UQ4	& 2	& 1 & 2 & 0 & 0 \\ \hline
	UQ5	& 3	& 1	& 2:1 & 0 & 0 \\ \hline
	UQ6	& 5	& 3	& 3:1:1 & 2 & 0 \\ \hline
	UQ7	& 5	& 3	& 3:1:1	& 2 & 0 \\ \hline
	UQ8	& 7	& 3	& 4:2:1 & 2	& 0 \\ \hline
	UQ9	& 5	& 2	& 3:2 &	1 &	0 \\ \hline
	UQ12 & 2 & 1 & 2 & 0 & 0 \\ \hline
	CRQ7 & 6 & 3 & 1:4:1 & 2 & 0 \\ \hline
    \end{tabular}	
    \quad    
    \begin{tabular}{| p{1.1cm} | p{0.72cm} | p{0.72cm} | p{1.31cm} | p{0.81cm} | p{0.81cm} | p{0.81cm} |}
    \hline
   	CRQ9 & 8 & 5 & 1:3:1:1:2 & 3 & 1 & \\ \hline
	CRQ13 & 4 & 3 &	1:2:1 &	2 &	0 & \\ \hline
	CRQ22 &	6 & 3 & 1:4:1 &	1 &	1 & \\ \hline
	CRQ23 & 7 & 4 & 2:1:2:2 & 2	& 1 & \\ \hline
	\textbf{Query (w/ $\cup$)} & \textbf{\#TP} & \textbf{\#STP} & \textbf{\#Edges in STP} & \textbf{\#S-O $\Join$} & \textbf{\#O-O $\Join$} & \textbf{\#Br in $\cup$} \\ \hline		
	UQ1+ & 17 &	17 & 1 & 0 & 0 & 17 \\ \hline
	UQ2+ & 51 & 20 & 3 & 0 &	0 & 17 \\ \hline
	UQ3	& 7	& 3	& (4:1)/(4:0):1 & 2 & 0 & 2 \\ \hline
	UQ4+ & 24 & 12 & 2 & 0 & 0 & 12 \\ \hline
	UQ12+ &	24 & 12 & 2 & 0 & 0 & 12 \\ \hline
	UQ18 & 10 & 5 & 2 & 5 &	0 &	6 \\ \hline
    \end{tabular}
    }
    \vspace {7pt}
	\caption{Characteristics of Testbed Queries}
	\label{QueryDetails}
	\vspace {-45pt}
	\end{table}
\subsection{Baseline Comparison - Non-UNION Queries}
\begin{figure}[h!]

\vspace {-18pt}
\centering
\includegraphics[width=\linewidth]{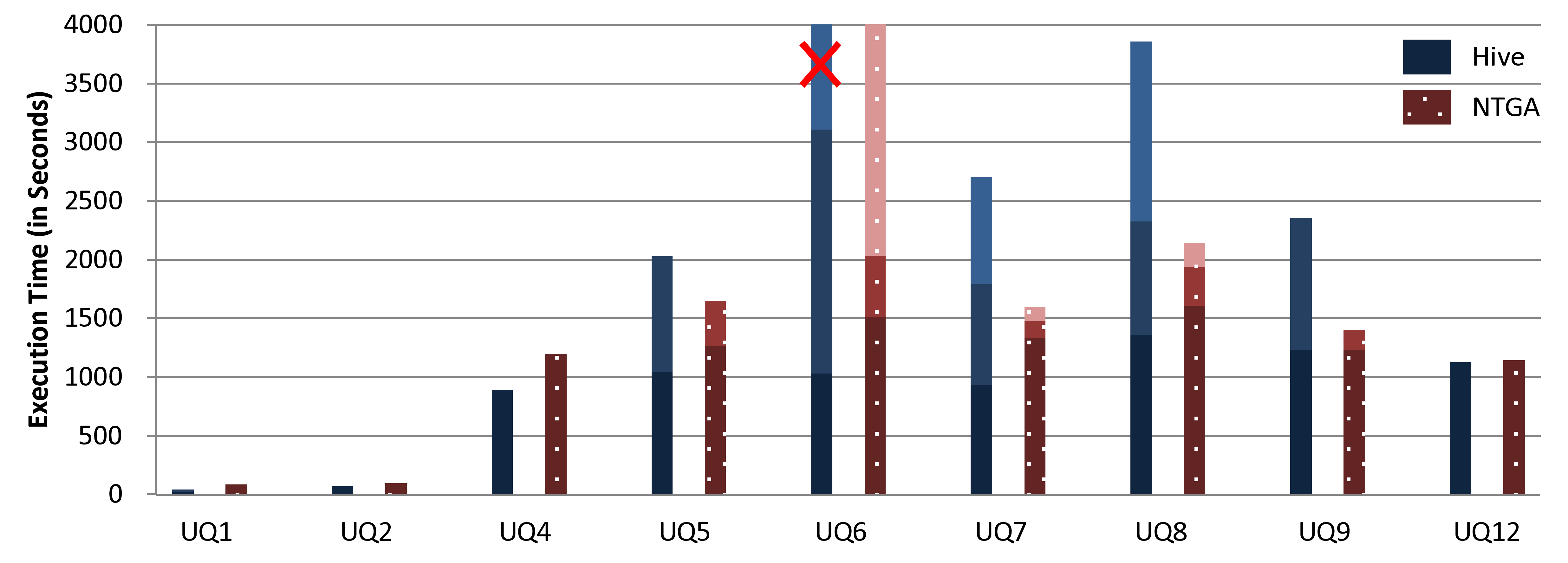}\vspace*{-1pt}
\caption{Execution Time of Non-union Queries on Uniprot}
\label{Eval1}
\vspace {-20pt}
\end{figure}

Fig.~\ref{Eval1} and Fig.~\ref{Eval2}(a) show the execution time of the base-line queries on Hive and NTGA. (Each different color block/shading in a bar represents a single MR job). NTGA-based plans often outperform relational ones, particularly when a query has multiple star patterns (\CodeIn{UQ6-UQ8}) and/or when star patterns are denser, i.e., contain more triple patterns (\CodeIn{UQ7} vs. \CodeIn{UQ8}). On the other hand, if a query contains only a single triple pattern or a star with very few triple patterns (\CodeIn{UQ1}, \CodeIn{UQ2}, \CodeIn{UQ4}, and \CodeIn{UQ12}), then NTGA-based execution plan has a slight performance penalty. For example, in query \CodeIn{UQ1}, NTGA took an additional 45 seconds because Hive processes such queries in a single map phase while NTGA requires both the map and reduce phases. For the cases where both NTGA and Hive use both phases of the single MR cycle (single star pattern, e.g., \CodeIn{UQ2}), Hive spends much less effort in the reduce phase simply assembling tuples that share the same join key while NTGA has the overhead of creating the data structures needed to manage triplegroups. However, as we see from the queries with multiple star patterns, this overhead is amortized over subsequent cycles and ultimately produces an advantage for NTGA as seen in the other queries. A similar performance pattern is observed for most Chem2Bio2RDF queries for similar reasons. However, here the performance advantage of NTGA over HiveQL plans is at a much smaller scale because the dataset is orders of magnitude smaller. In case of \CodeIn{UQ6}, Hive fails its execution due to the skewness producing very large intermediate result sizes. This is mainly caused from the join between two highly multi-valued properties, i.e. while most reducers for the join were able to be finished, remaining few reducers repeatedly failed due to the lack of memories. In case of NTGA, smaller footprints from its implicit data representation allows the execution of the query, avoiding the memory related issues. We expect that the performance of NTGA shown here can be further improved when some of the optimizations for dealing with MVPs from our previous works~\cite{SWIMRAPID} are integrated. 
\subsection{Union Queries}
\begin{figure}[h!]
\vspace {-18pt}
\centering
\includegraphics[width=\linewidth]{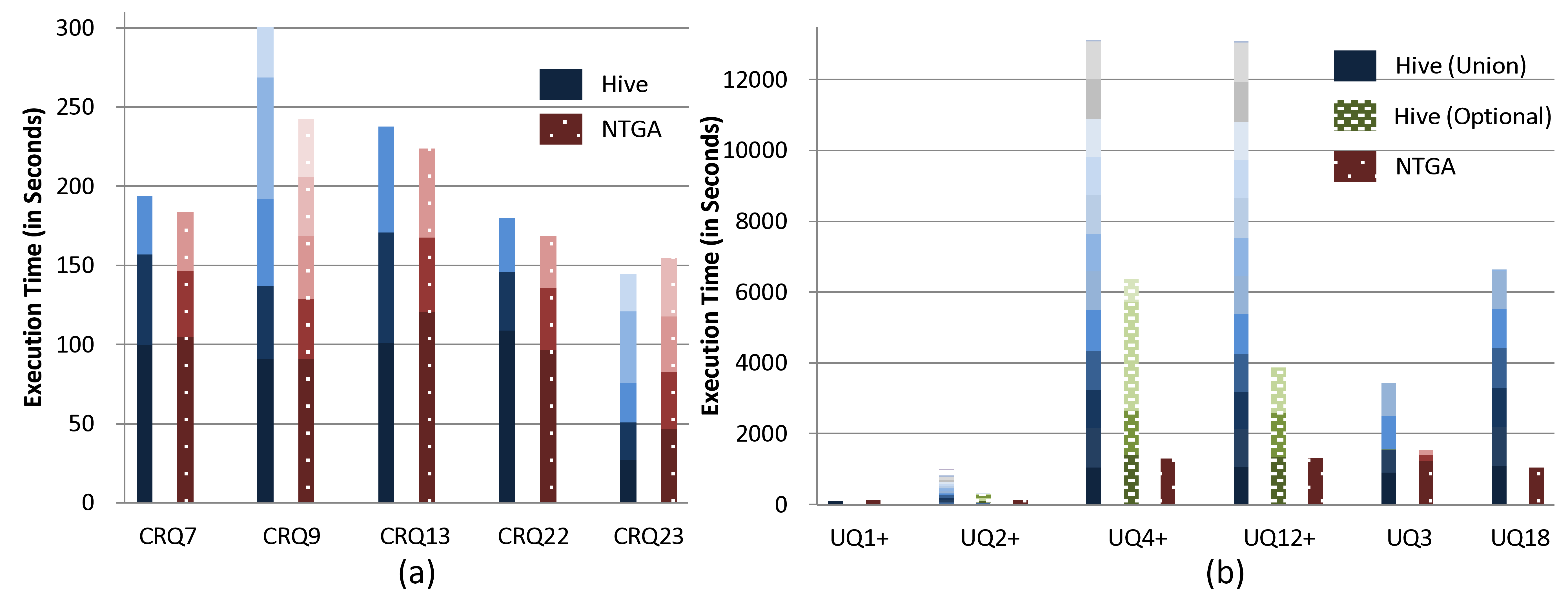}\vspace*{-1pt}
\caption{Execution Time of (a) Non-Union Queries on Chem2bio2RDF and (b) Union Queries on Uniprot}
\label{Eval2}
\vspace {-20pt}
\end{figure}

In general, Hive(Union) plans performed the worst out of the three approaches because the length of the MR execution workflow depends on the size of rewritings, i.e., the width(\# of branches) of the UNION clause. For example, the input files are read 12 times for \CodeIn{UQ4+} and \CodeIn{UQ12+} where each MR cycle is used to process one branch of the UNION. This is because Hive is unable to identify and exploit ``correlation'' across the UNION subqueries because they are not single triple pattern. Hive(Optional) plans produced using the MQO rewriting technique typically comprise of multiple $n$ left outer joins (LOJ) where often, several or all LOJ branches have star patterns that are correlated. This technique produces a superset of results, i.e., may contain false positives so that an additional MR cycle is needed to filter out incorrect matches. In general, Hive(Optional) plans showed better performance than the Hive(Union) plans. For example, Hive(Optional) showed 2.35 times better performance than Hive(Union) for \CodeIn{UQ12+} because its number of MR jobs reading input files is 1. Although Hive(Optional) may use more disk write phases than the Hive(Union) plan for some queries, e.g., \CodeIn{UQ4+} and \CodeIn{UQ12+} due to writing/filtering of false positives, the amount of data read was smaller. Hive groups the execution of the correlated star subqueries across the OPTIONAL branches, in this case into 3 groups each processing $n/3$ join statements. On the other hand, the NTGA execution plan is able to merge all such correlated subqueries into a single execution cycle allowing input files to be scanned only once. This allows NTGA to outperform the other two approaches for the union graph patterns with large widths($\ge 5$). For example, Fig.~\ref{Eval2}(b) shows that for \CodeIn{UQ12+}, NTGA was approximately 9 and 4 times faster than Hive(Union) and Hive(Optional), respectively. 

%% file: relatedwork.tex
\section{Related Work}
There have been an extensive effort for efficient and flexible querying of divergent life science datasets, e.g., a selection of proper query sources~\cite{BioGuide}, a central portal system~\cite{guberman2011biomart}, and a federation approach~\cite{cheung2009journey}. Processing SPARQL over large amounts of RDF data on MapReduce has been actively investigated (e.g., \cite{HuangLargeSPARQL,HadoopRDF}), but optimizing disjunctive queries was not discussed and life science datasets were rarely evaluated. As an optimization for union queries, union pushdown~\cite{UnionPushDown} technique is introduced to optimize disjunctive queries with steps utilizing hybrid methods of CNF-/DNF-based optimizations in relational database. The main difference with our approach is a that the union-pushdown does not remove UNION operators, but push down under other common operators so that the UNION operators accept different (star-)join operators as operands. In NTGA-based plan, the UNION operators are implicitly removed and represented as disjunctive ORs in the parameter of \CodeIn{TG\_GroupFilter} and \CodeIn{TG\_UJoin}.

\section{Conclusion and Future Work}
In this paper, we presented a comparative discussion on evaluating union graph pattern queries on MapReduce using relational-style algebra optimizations and optimizations based on a different algebra, NTGA. Such queries are important for querying heterogeneous datasets common in the life sciences domain. We presented a performance evaluation using two real-world life sciences datasets and queries which showed the superiority of the NTGA algebra. We also highlighted some future work issues for example model for better estimation of resources needed for NTGA plans.

%% file: main.bbl
\begin{thebibliography}{10}

\bibitem{IEEEMagazine}
Anyanwu, K., Kim, H., Ravindra, P.:
\newblock {Algebraic Optimization for Processing Graph Pattern Queries in the
  Cloud}.
\newblock Internet Computing, IEEE \textbf{17}(2) (2013)  52--61

\bibitem{Uniprot}
Apweiler, R., Bairoch, A., Wu, C.H., Barker, W.C., Boeckmann, B., Ferro, S.,
  Gasteiger, E., Huang, H., Lopez, R., Magrane, M.,  et~al.:
\newblock {UniProt: the Universal Protein Knowledgebase}.
\newblock Nucleic Acids Research \textbf{32}(suppl 1) (2004)  D115--D119

\bibitem{UnionPushDown}
Chang, J.Y., goo Lee, S.:
\newblock {An optimization of disjunctive queries: union-pushdown}.
\newblock In: Proc. COMPSAC. (1997)  356--361

\bibitem{CHEM2BIO2RDF}
Chen, B., Ding, Y., Wang, H., Wild, D., Dong, X., Sun, Y., Zhu, Q.,
  Sankaranarayanan, M.:
\newblock {Chem2Bio2RDF: A Linked Open Data Portal for Systems Chemical
  Biology}.
\newblock In: {Proc. WI-IAT}. Volume~1. (2010)  232--239

\bibitem{cheung2009journey}
Cheung, K.H., Frost, H.R., Marshall, M.S., Prud'hommeaux, E., Samwald, M.,
  Zhao, J., Paschke, A.:
\newblock {A Journey to Semantic Web Query Federation in the Life Sciences}.
\newblock BMC bioinformatics \textbf{10}(Suppl 10) (2009)  S10

\bibitem{BioGuide}
Cohen-Boulakia, S., Davidson, S., Froidevaux, C.:
\newblock {A User-Centric Framework for Accessing Biological Sources and
  Tools}.
\newblock In: Data Integration in the Life Sciences. Volume 3615.
\newblock 0 (2005)  3--18

\bibitem{MapReduce}
Dean, J., Ghemawat, S.:
\newblock {MapReduce: Simplified Data Processing on Large Clusters}.
\newblock Commun. ACM \textbf{51}(1) (January 2008)  107--113

\bibitem{demir2010biopax}
Demir, E., Cary, M.P., Paley, S., Fukuda, K., Lemer, C., Vastrik, I., Wu, G.,
  D'Eustachio, P., Schaefer, C., Luciano, J.,  et~al.:
\newblock {The BioPAX Community Standard for Pathway Data Sharing}.
\newblock Nature biotechnology \textbf{28}(9) (2010)  935--942

\bibitem{Gottlob}
Gottlob, G., Orsi, G., Pieris, A.:
\newblock {Ontological Queries: Rewriting and Optimization}.
\newblock In: {Proc. ICDE}. (2011)  2--13

\bibitem{guberman2011biomart}
Guberman, J.M., Ai, J., Arnaiz, O., Baran, J., Blake, A., Baldock, R., Chelala,
  C., Croft, D., Cros, A., Cutts, R.J.,  et~al.:
\newblock {BioMart Central Portal: an open database network for the biological
  community}.
\newblock Database: the journal of biological databases and curation
  \textbf{2011} (2011)

\bibitem{HuangLargeSPARQL}
Huang, J., Abadi, D.J., Ren, K.:
\newblock {Scalable SPARQL Querying of Large RDF Graphs}.
\newblock Proc. {VLDB} \textbf{4}(11) (2011) ~0

\bibitem{HadoopRDF}
Husain, M., McGlothlin, J., Masud, M., Khan, L., Thuraisingham, B.:
\newblock {Heuristics-Based Query Processing for Large RDF Graphs Using Cloud
  Computing}.
\newblock \textbf{23}(9) (2011)  1312--1327

\bibitem{MQO}
Le, W., Kementsietsidis, A., Duan, S., Li, F.:
\newblock {Scalable Multi-query Optimization for SPARQL}.
\newblock In: {Proc. ICDE}. (2012)  666--677

\bibitem{ESWCRAPID}
Ravindra, P., Kim, H., Anyanwu, K.:
\newblock {An Intermediate Algebra for Optimizing RDF Graph Pattern Matching on
  MapReduce}.
\newblock In: Proc. ESWC. (2011)  46--61

\bibitem{SWIMRAPID}
Ravindra, P., Kim, H., Anyanwu, K.:
\newblock {To Nest or Not to Nest, When and How Much: Representing Intermediate
  Results of Graph Pattern Queries in MapReduce-based Processing}.
\newblock In: Proc. SWIM. (2012)  5:1--5:8

\bibitem{TradeOff}
Stuckenschmidt, H., Broekstra, J.:
\newblock {Time-Space Trade-offs in Scaling up RDF Schema Reasoning}.
\newblock In: {Proc. WISE}. (2005)  172--181

\end{thebibliography}
